\documentclass[fleqn,12pt,twoside]{article}
\usepackage{espcrc1}

\usepackage{graphicx}
\usepackage{epsfig}
\usepackage[figuresright]{rotating}

\newcommand{\be}{\begin{equation}}
\newcommand{\en}{\end{equation}}
\newcommand{\bea}{\begin{eqnarray}}
\newcommand{\ena}{\end{eqnarray}}

\newcommand{\hbo}{\hbox to 1 true cm {\hfill } }
\newcommand{\tr}{\hbox{tr}}

\newcommand{\AmS}{{\protect\the\textfont2
  A\kern-.1667em\lower.5ex\hbox{M}\kern-.125emS}}

\title{Confinement versus color superconductivity: 
a lattice investigation }

\author{Kurt Langfeld\address[ITU]{Institut f\"ur Theoretische Physik, 
        Universit\"at T\"ubingen, Auf der Morgenstelle 14, \\ 
        72076 T\"ubingen, Germany }
        \thanks{I thank E.-M.~Ilgenfritz, S.~Hands and H.~Reinhardt for 
         helpful dicussions.},
 }

\begin{document}

% typeset front matter
\maketitle

\begin{figure}[t]
\centerline{
\epsfxsize=0.4\linewidth 
\epsfbox{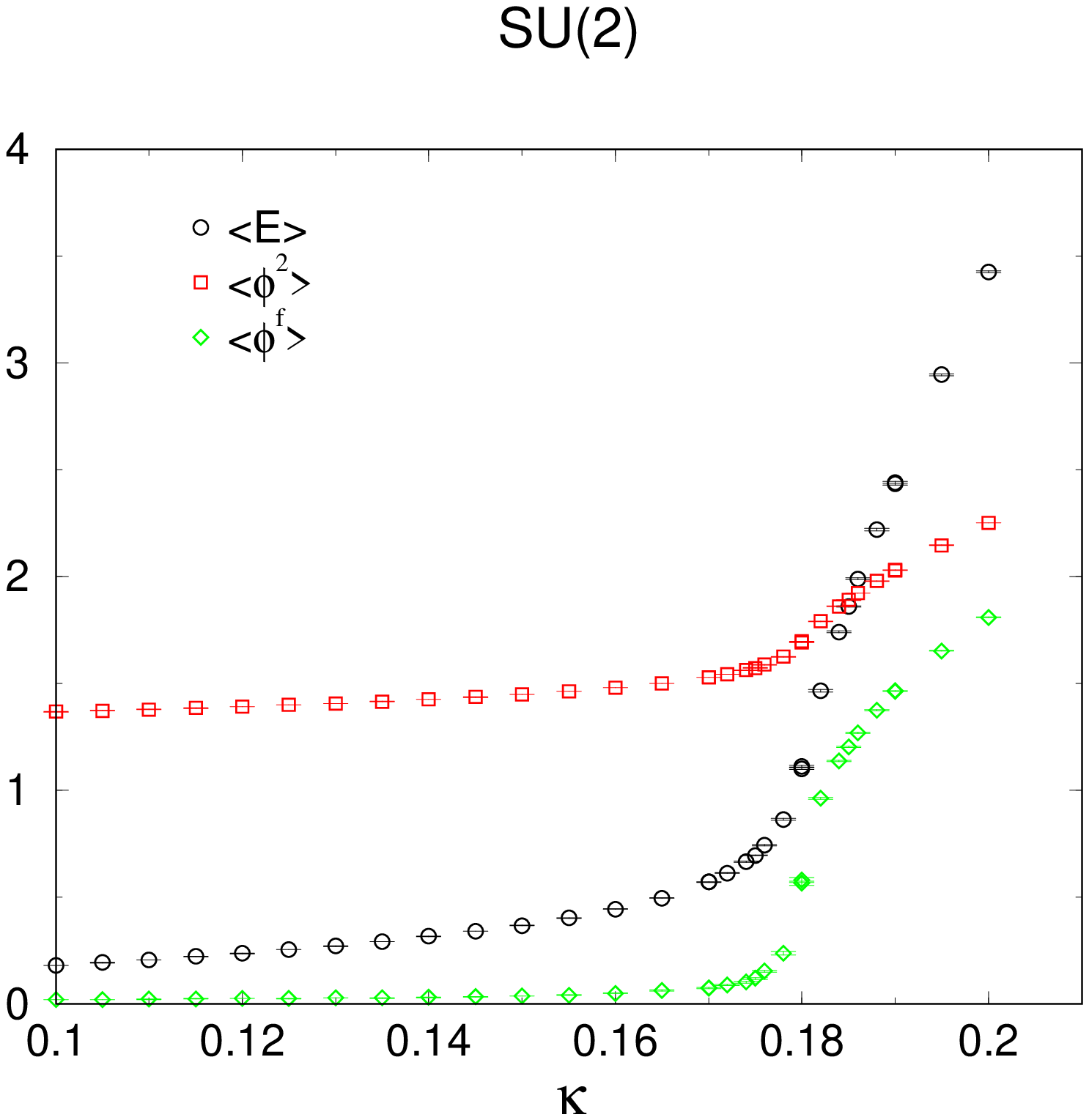}
\epsfxsize=0.44\linewidth 
\epsfbox{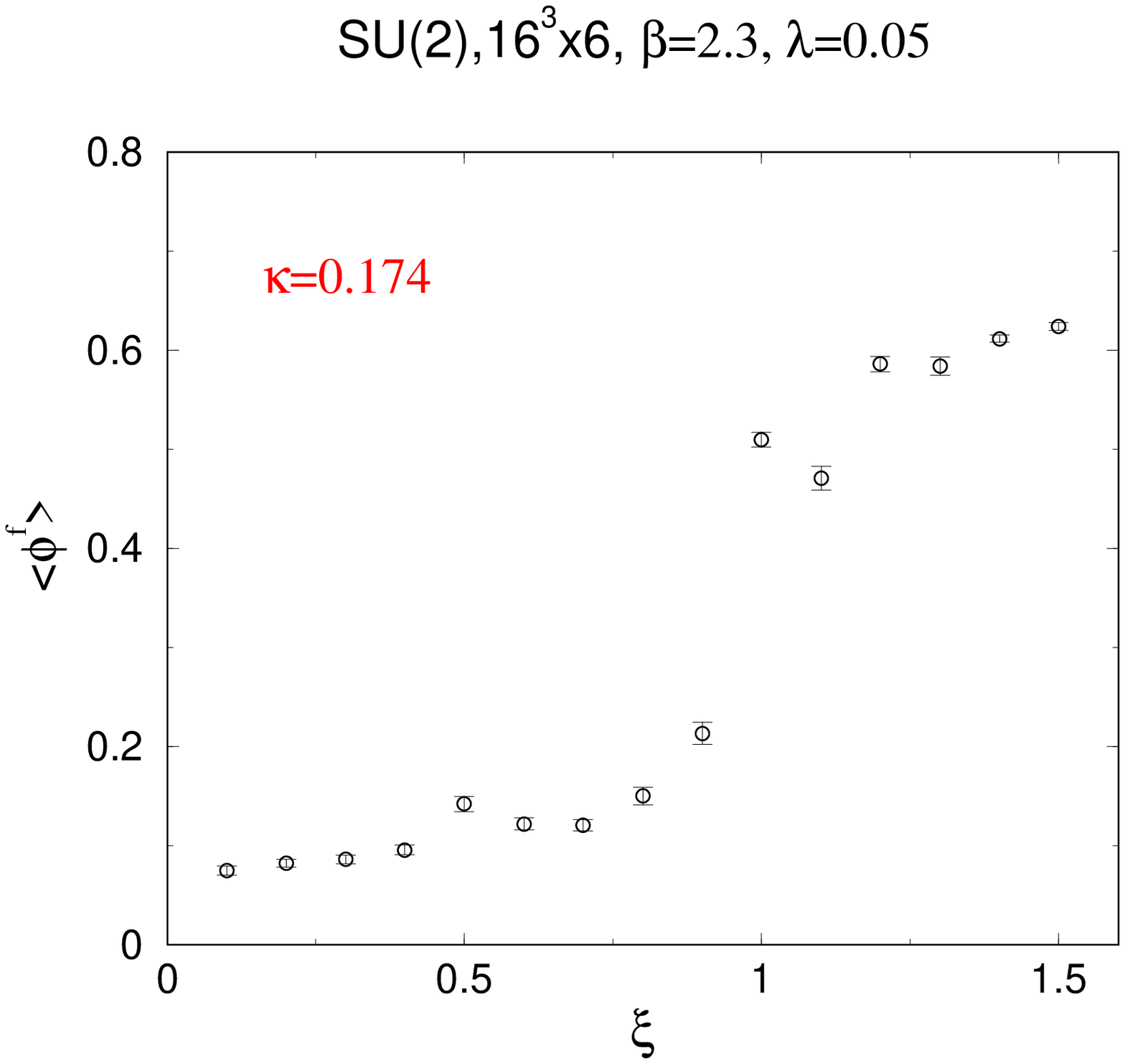}
}
%\caption{xxx}.
\label{fig:1}
\end{figure}

The QCD phase diagram as function of the temperature and the baryon 
chemical potential is the object of many theoretical and experimental 
investigations. At a high chemical potential, QCD is in the deconfined phase 
with quarks forming a Fermi surface. It turns out that the gluon induced 
quark interaction is attractive in the diquark channel~\cite{bai94} 
generating a Cooper instability of the Fermi surface and leading 
to {\it color superconductivity}~\cite{csc} (CSC). 
At the critical value of the chemical potential, the confining forces 
turn the Fermi surface of quarks into a Fermi surface of hadrons. 
Several proposals for a description of matter close to the confinement 
regime have been made: a chiral 
restored phase might occur in which vector mesons are the light 
excitations~\cite{la97b}. Alternatively, it might appear that 
the hadrons smoothly convert into quarks moving in a diquark 
condensate~\cite{son01,wil99}.

\vskip 0.2cm 
The aim of the present project is to propose an effective model which 
is able to cope with confinement on one hand and to describe finite 
density effects on the other hand. 

\vskip 0.2cm 
Here, I will use the lattice approach of~\cite{la00} in the leading order 
of the heavy quark approximation. A scalar auxiliary field $\phi $ 
is supplemented to the model which mimics Fermi surface effects 
which are not contained in the leading order mass approximation. 
Since the diquark field belongs to the $[\bar{3}]$ representation, 
it possesses the quantum numbers of a scalar field in the fundamental 
representation and is identified with the auxiliary Higgs field. Close to 
the deconfinement phase transition, I neglect the substructure of 
the diquarks, and propose the following  effective lattice action to describe 
the confinement CSC transition:
\begin{eqnarray}
S_{latt} &=& \sum _{\{x\}} \biggl( \beta  \, \sum _{\mu > \nu } 
P_{\mu \nu }(x) \; + \; \chi \, 
\left[ e^{ \frac{\mu - m }{T} } \; \tr P(x) \; + \; 
e^{ - \frac{\mu + m }{T} } \; \tr P^\dagger (x) \, 
\right] \; \biggr) \; + \; S_h \; , 
\label{eq:1} \\ 
S_h &=& \kappa \; \sum _{\{x\}, \mu } \; \tr \phi ^\dagger _x 
U_{x,\mu } \phi _{x+\mu} \; + \; \sum _{\{x\}} \left[ 
\phi^\dagger _x \phi _x \; + \; \lambda \left( \phi^\dagger _x \phi _x 
-1 \right)^2 \right] \; , 
\label{eq:2} 
\end{eqnarray} 
where $U_{x,\mu }$ is the link variable, $P_{\mu \nu }(x)$ is the plaquette 
and $\phi_x$ comprises the four component Higgs field acting as 
the collective diquark field. Color superconductivity is related to 
the Higgs phase of the model. $P(x)$ is the Polyakov line starting at 
space-time point $x$ and wrapping around the torus in time direction.
$m$ is the (heavy) quark mass, and $\chi $ depends on the 
temperature and $m$ 
(for details see~\cite{la00}). The parameter $\beta $ is related to the 
gauge coupling as usual. The parameters 
$\lambda $ and $\kappa $ describe the properties of the collective diquark 
field. The fact that the design of the effective diquark 
theory reflects the QCD symmetries stirs the hope that already the 
SU(2) simulations provide a qualitative understanding of the 
transition region from confinement to CSC. 

\vskip 0.2cm 
For a certain choice of the parameters, i.e. $\lambda = 0.0017235$, 
$\chi =0$, 
$\beta =8$, $\kappa _c \approx 0.12996 $, the above model describes 
the electro-weak phase transition as e.g.~signalled by the Higgs 
hopping term, i.e~$\langle E \rangle = \langle \tr \phi ^\dagger _x 
U_{x,\mu } \phi _{x+\mu } \rangle $. 

\vskip 0.2cm 
Here, I advertise that the Higgs phase is characterized by a 
spontaneous breakdown of the global color symmetry which is a leftover of 
gauge fixing. Here, I used the familiar  Landau gauge 
$ 
\sum _{\{x\},\mu } \tr \; \Omega (x) \, U_\mu (x) \, \Omega ^\dagger 
(x+\mu) \; \stackrel{\Omega }{\rightarrow } \; \mathrm{max} \; . 
$ 
For this task, an improved simulated annealing algorithm was employed 
(details will be presented elsewhere). If $\phi ^f$ denotes the gauge 
fixed Higgs field, I define 
$ 
\langle \phi ^f \rangle \; = \; \left\langle \left( 
\frac{1}{N} \sum _x \phi ^f(x) \right)^2 \right\rangle ^{1/2} \; . 
$
If the residual 
global color symmetry is realized, one finds $\sum _x \phi ^f(x)=0$. 
A spontaneous breakdown of this symmetry would allow 
for $\langle \phi ^f \rangle \not= 0$. My lattice results (see figure 
right panel) nicely show that the Higgs regime is indeed characterized
by $\langle \phi ^f \rangle \not=0$. 

\vskip 0.2cm 
Choosing the parameter set $\lambda = 0.05$, $\beta =2.3$, 
$16^3 \times 6 $, the transition point at $\kappa _c \approx 0.18 $
distinguishes 
between the confinement and the CSC phase. Choosing $\kappa = 0.174$, the 
vacuum is in the confinement phase. I observe that increasing 
the chemical potential induces the transition to the CSC phase 
(see figure right panel, $\xi = \chi \, \exp \{ (\mu - m )/T \}$).

\end{document}